\documentclass[reprint, aip, amsmath, groupedaddress]{revtex4-1}
\usepackage[mathlines]{lineno}
\modulolinenumbers[5]
\usepackage{graphicx} 
\usepackage[hidelinks]{hyperref}

\begin{document}


\title[Dose reconstruction through deconvolution of dosimeter responses]{Technical Note: Reconstruction of physical and biological dose distributions of carbon-ion beam through deconvolution of longitudinal dosimeter responses}

\author{Nobuyuki Kanematsu}
\email{kanematsu.nobuyuki@qst.go.jp}
\homepage[\\]{https://orcid.org/0000-0002-2534-9933}

\author{Taku Inaniwa}
\author{Shunsuke Yonai}

\author{Hideyuki Mizuno}

\affiliation{National Institute of Radiological Sciences, QST; 4-9-1 Anagawa, Inage-ku, Chiba 263-8555, Japan}

\date{\today}

\begin{abstract}
\begin{description}
\item[Purpose]
This is a theoretical simulation study for proof of concept of radiochromic film dosimetry to measure physical and biological doses without plan-based quenching correction for patient-specific quality assurance of carbon-ion radiotherapy.
\item[Methods]
We took a layer-stacking carbon-ion beam comprised of range-shifted beamlets.
The dosimeter response was simulated according to an experimental quenching model.
The beam model followed a treatment planning system.
The beam was decomposed into finely arranged beamlets with weights estimated by deconvolution of longitudinal dosimeter responses.
The distributions of physical and biological doses were reconstructed from the estimated weights, and were compared with the plan.
We also evaluated the sensitivity to measurement errors and to erratic delivery with an undelivered beamlet. 
\item[Results]
The reconstructed physical and biological doses accurately reproduced the simulated delivery with errors approximately corresponding to the measurement errors. 
The erratic beam delivery was easily detectable by comparison of biological dose distribution to the plan.
\item[Conclusions]
We have developed a method to measure physical and biological doses by longitudinal dosimetry of quenched response without using plan data.
The method only involves a general optimization algorithm, a radiobiology model, and experimental beamlet data, and requires no extra corrections.
Theoretically, this approach is applicable to various dosimeters and to proton and ion beams of any delivery method, regardless of quenching or biological effectiveness.
\end{description}
\end{abstract}


\maketitle

\section{Introduction}

In state-of-the-art radiotherapy (RT), highly complex beams are delivered for optimum treatment of individual patients, for which quality assurance (QA) must always be secured.
In patient-specific QA, the quality of treatment beams is verified by dose comparison between measurements and calculations of a QA plan with a phantom in place of the patient before treatment.\cite{Miften2018}
On such occasions, self-developing radiochromic films (RCF), such as Gafchromic\textsuperscript{\texttrademark} EBT3 (Ashland Inc., Bridgewater, NJ), are conveniently used for high-resolution planar dosimetry.

For proton and ion beams, however, RCF dosimetry suffers from quenching, namely sensitivity reduction with linear energy transfer (LET) of ionizing particles.\cite{Battaglia2016,Castriconi2016}
The quenching effect is especially unfavorable in carbon-ion RT,\cite{Kamada2015} which utilizes its high-LET Bragg peak to enhance the relative biological effectiveness (RBE) for tumor cells.\cite{Kanai1999}
The dependence of RCF sensitivity on dose-averaged LET has been evaluated experimentally with clinical carbon-ion beams for potential application to quenching correction in dosimetry.\cite{Yonai2018}
Such a correction will require prior knowledge of the spatial LET distribution, which is not generally measurable in a clinical setup and should be obtained from a treatment plan somehow.\cite{Kanematsu2018}
The involvement of plan-based correction poses a question for the validity of dosimetric QA as an independent verification against the plan, especially because LET and dose, namely energy transfer and absorption, are intrinsically correlated.
Alternatively, patient-specific QA may be achievable by direct comparison of dosimeter responses between measurement and calculation by a convolution algorithm configured with a special kernel for the response.\cite{Miyatake2013}

In carbon-ion RT, clinical assessment of treatment plans is generally performed with distributions of RBE-weighted dose or biological dose, which is normally uniform over a spread-out Bragg peak (SOBP).
In patient-specific QA, the biological dose distribution is naturally desired for clinical assessment.
For a broad-beam delivery system with range modulation fixed per ridge filter, the longitudinal physical dose profiles measured with a multilayer ionization chamber have been successfully converted to biological dose profiles based on the predefined beam data for each modulation.\cite{Mizota2002}
That approach is, however, not applicable to variable range modulation planned specifically for each treatment with scanning or layer-stacking beams.\cite{Haberer1993, Kanematsu2002}

In this study, we propose a new dosimetric method to measure physical and biological doses without plan-based correction for QA of carbon-ion RT.
For this purpose, we conceptually follow the longitudinal RCF dosimetry proposed for proton beams,\cite{Zhao2010} apply the deconvolution-and-reconstruction approach proposed for in vivo dosimetry with positron emission tomography after proton RT,\cite{Fourkal2009} and undertake a proof of concept through theoretical simulation.

\section{Materials and Methods}

\subsection{Simulation apparatus}

\subsubsection{Carbon-ion beam}

\begin{figure}
\includegraphics{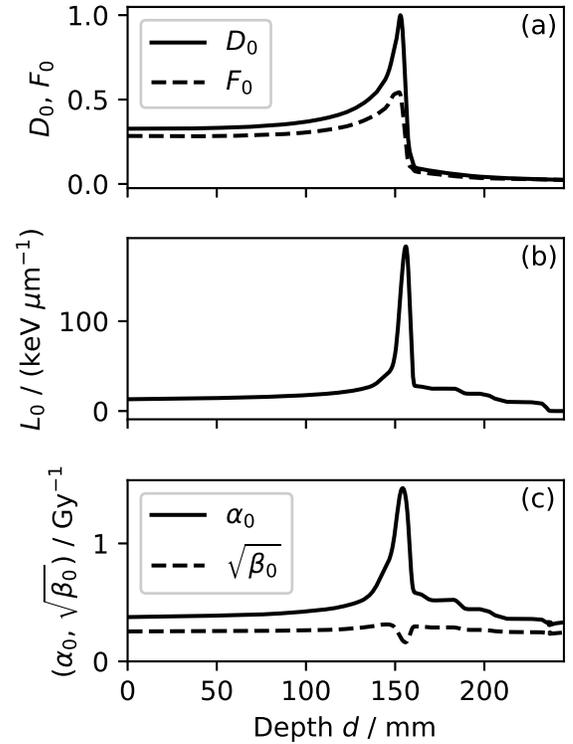}
\caption{(a) Relative dose $D_0$ with modeled RCF response $F_0$, (b) dose-averaged LET $L_0$, and (c) dose-averaged LQ parameters $\alpha_0$ and $\sqrt{\beta_0}$ for a human salivary gland tumor cell line as functions of depth in water $d$ for a ripple-filtered carbon-ion beam of 290 MeV per nucleon, configured in a treatment planning system.\cite{Kanematsu2002}}
\label{fig1}
\end{figure}

We took a broad carbon-ion beam of 290 MeV per nucleon moderated by a Gaussian ripple filter ($\sigma=1.8$ mm) having an available beam range of 154.2 mm in water,\cite{Kanematsu2002} for an elementary beam or beamlet.
Figure~\ref{fig1} shows dose $D_0$, dose-averaged LET $L_0$, and dose-averaged radiosensitivity parameters $\alpha_0$ and $\sqrt\beta_0$ of the linear-quadratic (LQ) model as functions of depth, for the basis beamlet, namely the one without range shifting.
A beam modulated by layer stacking is comprised of range-shifted beamlets in dose,
\begin{eqnarray}
&& D_\textrm{M}(d) = \sum_i w_i \, D_i(d)
\quad\text{with}\\
&& D_i(d) = D_0(d+s_i),
\end{eqnarray}
where $d$ is the depth in water and $s_i$ and $w_i$ are the range shift and the weight of beamlet $i$.

\subsubsection{Dosimeter response}

To simulate RCF responses, we applied an empirical model by Yonai et al.\cite{Yonai2018} 
In their dosimetry with Gafchromic\textsuperscript{\texttrademark} EBT3 calibrated for a 6-MV x ray, the relative efficiency $\epsilon$ of dose response to carbon-ion beams was formulated as a function of dose-averaged LET $L$,
\begin{equation}
\epsilon(L) = 1- 0.0732 \, \left(\frac{\mathrm{\mu m}}{\mathrm{keV}} L - 8.65\right)^{0.388},
\end{equation}
for $L > 8.65$ keV/$\mu$m or $\epsilon = 1$ otherwise.
Accordingly, dose $D$ would be measured as dosimeter response $F = \epsilon D$. 
Assuming the additivity, we can decompose a modulated beam into beamlets in the dosimeter response,
\begin{eqnarray}
&& F_\textrm{M}(d) = \sum_i w_i\, F_i(d)
\quad\text{with}\\
&& F_i(d) = F_0(d + s_i)
\quad\text{and}\quad
 F_0 = \epsilon(L_0) D_0,
\end{eqnarray}
where $F_i$ is the dosimeter response to beamlet $i$ and $F_0$ is that to the basis beamlet, which is also shown in Fig.~\ref{fig1}(a).

\subsubsection{Biological response}

The radiation mixing due to the modulation was handled by dose-weighted averaging of radiosensitivity parameters $\alpha_i(d) = \alpha_0(d+s_i)$ and $\sqrt{\beta_i}(d) = \sqrt{\beta_0}(d+s_i)$ over the beamlets,\cite{Zaider1980}
\begin{eqnarray}
&& \alpha_\mathrm{M} = \frac{\sum_i w_i \, D_i\, \alpha_i}{D_\mathrm{M}}
\\
&& \sqrt{\beta}_\mathrm{M} = \frac{\sum_i w_i \, D_i\, \sqrt{\beta_i}}{D_\mathrm{M}}.
\end{eqnarray}
The biological dose $B_\mathrm{M}$, formulated as
\begin{equation}
B_\mathrm{M} =\sqrt{\frac{\alpha_\mathrm{x}^2}{4 \beta_\mathrm{x}^2} + \frac{\alpha_\mathrm{M} D_\mathrm{M} + \beta_\mathrm{M} D_\mathrm{M}^2} {\beta_\mathrm{x}}} - \frac{\alpha_\mathrm{x}}{2 \beta_\mathrm{x}},
\end{equation}
is the dose of a reference x-ray radiation for a given survival fraction of reference cells, for which we took human salivary gland tumor cells with x-ray sensitivity parameters $\alpha_\mathrm{x} = 0.331$ Gy$^{-1}$ and $\beta_\mathrm{x} = 0.0593$ Gy$^{-2}$.\cite{Matsufuji2007}
The survival fraction of the reference cells $S$ is simply an exponential LQ function of biological dose $B$, with which the biological effect is defined as
\begin{equation}
E(B) = –\ln S(B) = \alpha_\mathrm{x} B + \beta_\mathrm{x} B^2.
\end{equation}

\subsubsection{Range modulation}

We arranged 32 beamlets with range shifts of 4.2 mm to 81.7 mm at 2.5-mm intervals to form a modulated beam having a SOBP size of 80 mm and a beam range of 150 mm in water.
With decomposing the biological effect into beamlet contributions,
\begin{eqnarray}
&& E_\mathrm{M} = \alpha_\mathrm{M} D_\mathrm{M} + \beta_\mathrm{M} D_\mathrm{M}^2 = \sum_i w_i E_i
\quad\text{with}\\
&& E_i = \left( \alpha_i + \sqrt{\beta_i} \sqrt{\beta_\mathrm{M}} D_\mathrm{M} \right) D_i,
\end{eqnarray}
we optimized the beamlet weights to give a uniform biological effect in the SOBP by convergence of multiplicative recursion,\cite{Lomax1999}
\begin{equation}
w_i := \frac{\sum_j E_i^2(d_j) \, E(B)} {\sum_j E_i^2(d_j) \, E_\mathrm{M}(d_j)} w_i, 
\end{equation}
where $d_j$ is a SOBP sampling depth, for which we took the peak-dose depth of each beamlet, and we prescribed $B = 4.04$~Gy for 10\% survival.

\subsection{QA dosimetry simulation}

\subsubsection{Measurement}

For the modulated beam, we simulated longitudinal RCF dosimetry for depths from 0 mm to 200 mm sampled at 1-mm intervals with random measurement errors of a normal distribution, and obtained the dosimeter response at each sampling depth $d_j$ by
\begin{eqnarray}
\hat{F}_j = (1 + \sigma_\mathrm{N} \, {r_\mathrm{N}}_j)F_\textrm{M}(d_j),
\end{eqnarray}
where ${r_\mathrm{N}}_j$ is a random number of the standard normal distribution and $\sigma_\mathrm{N}$ is a relative standard uncertainty of the response measurement.

\subsubsection{Deconvolution and reconstruction}

To decompose the modulation, we newly arranged possible beamlets of range shift $\hat{s}_i$ from 0 mm to the peak-dose depth of the basis beamlet at 0.5-mm intervals, and estimated their weights $\hat{w}_i$ by convergence of multiplicative recursion,
\begin{eqnarray}
&& \hat{w}_i := \frac{\sum_j F_i^2(d_j) \, \hat{F}_j}{\sum_j F_i^2(d_j) \, F_\mathrm{R}(d_j)} \hat{w}_i 
\quad\text{with}\\
&& F_\mathrm{R}(d_j) = \sum_i \hat{w}_i F_i(d_j),
\end{eqnarray}
where $F_i$ is the dosimeter response to beamlet $i$ and $F_\mathrm{R}$ is the reconstructed dosimeter response. 
On each recursion, the minimum-weight beamlet was deleted if its weight was less than 1\% of the maximum, to eliminate insignificant beamlets.
Using the resultant beamlets, we also reconstructed the distributions of physical dose $D_\mathrm{R}$ and biological dose $B_\mathrm{R}$, and compared them with the plan.

\subsubsection{Sensitivity to errors}

We evaluated the sensitivity of the dosimetry to measurement errors by varying the relative standard uncertainty of response among $\sigma_\mathrm{N} =$ 0\%, 1\%, and 2\%, which may cover the dose uniformity of $\pm 2\%$ specified for EBT3 by the manufacturer. 
In addition, we intentionally deleted the 16th beamlet to test the detectability of such an erratic delivery of a corrupt plan with random errors of $\sigma_\mathrm{N} =$ 1\%.


\section{Results}

\begin{figure*}
\includegraphics{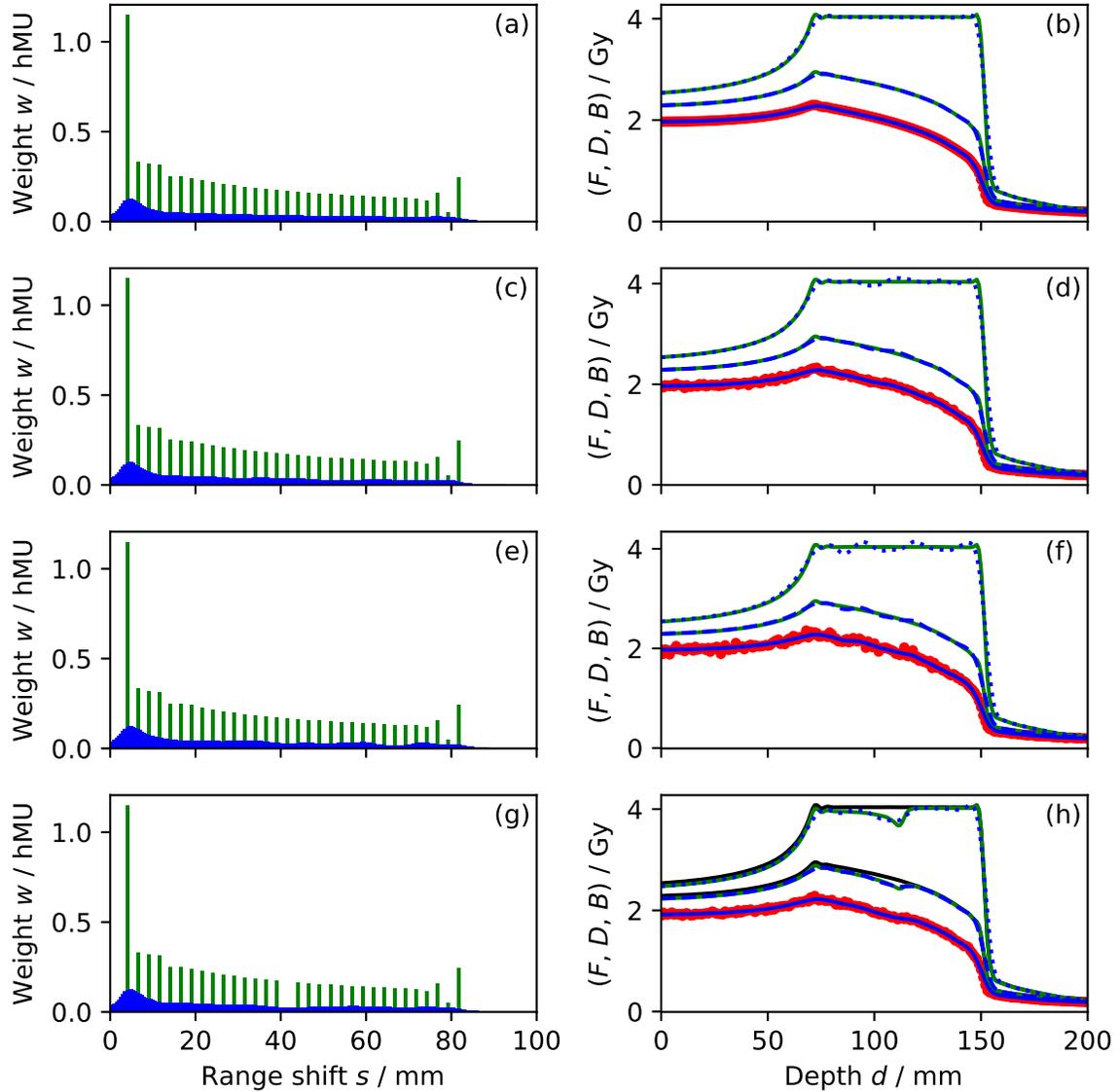}
\caption{QA dosimetry simulation for a layer-stacking beam of 150-mm range and 80-mm SOBP: (a) range-shift spectra of beamlet weights in hMU (or beamlet peak doses in Gy) for the modulations delivered ($w_i$, green) and estimated ($\hat{w}_i$, blue); (b) dosimeter responses measured ($F_j$, red markers) and reconstructed ($F_\mathrm{R}$, solid blue), physical doses delivered ($D_\mathrm{M}$, green) and reconstructed ($D_\mathrm{R}$, dashed blue), and biological doses delivered ($B_\mathrm{M}$, green) and reconstructed ($B_\mathrm{R}$, dotted blue); (c, d) and (e, f) are the same as (a, b) but with random measurement error of 1\% and 2\%, respectively; (g, h) is the same as (c, d) but with an undelivered beamlet and shown with planned doses (black).}
\label{fig2}
\end{figure*}

Figure~\ref{fig2} shows the range-shift modulations delivered and reconstructed, the dosimeter responses measured and reconstructed, and the physical and biological doses delivered and reconstructed in the QA dosimetry simulation.
The estimated weights did not exactly reproduce the delivered weights, even considering a factor of 5 difference in resolution.
As each beamlet had mixed ranges with a variation of a few mm by the ripple filter and natural straggling, the estimation was intrinsically limited in range-shift resolution.
Nevertheless, except at both ends of the SOBP, the reconstructed doses accurately reproduced the delivered doses.
In the SOBP, the dosimetric errors approximately corresponded to the random measurement errors.
The erratic beam delivery with an undelivered beamlet was easily detectable when compared with the plan.
In these plots, the errors were more apparent in biological dose than in physical dose having inherent gradient in the SOBP.

\section{Discussion}

We have demonstrated the theoretical feasibility of QA of layer-stacking carbon-ion RT by RCF dosimetry without using plan data.
The response precision of $\sigma_\mathrm{N}= 1\%$, which may be consistent with the EBT3 specification, would be sufficient for the standard QA tolerance of 2\% or 3\%.\cite{Miften2018}
The quenching simulation based on dose-averaged LET, originally for a carbon-ion beam of 60-mm SOBP,\cite{Yonai2018} could have been inaccurate for the ripple-filtered beam due to differences in the LET spectrum.
In reality, however, the quenched dosimeter response to the basis beamlet $F_0(d)$ should always be measured experimentally.
The assumption of additivity of RCF responses is supported by the fact that RCF is widely used for multifield and dynamic beam deliveries of various dose rates.\cite{Miften2018}

The dose distributions on the phantom measurement axis can further be mapped through water-equivalent depth onto the patient image for clinical evaluation.\cite{Mizota2002}
In addition, a film naturally offers two-dimensional dosimetry on multiple axes along the beam, and stacked films will offer three-dimensional dosimetry.
By implementing the radiobiology model that is common to treatment planning, QA practitioners can concentrate on beam-delivery errors.
In addition, they can also measure the biological dose in any radiobiology model by configuring the dosimetry system as such, which will be useful for the intercomparison of prescribed doses in a multi-institutional clinical research of carbon-ion RT with institutional model variations.\cite{Molinelli2016}

The deconvolution procedure replicates a common SOBP optimization, and is naturally applicable also to scanning and conventional broad beams.
To resolve range modulation, the use of a pristine Bragg peak for the basis beamlet would be ideal for its minimized range mixing, but it could also worsen random errors.
For energy scanning,\cite{Haberer1993} the acceleration energies, instead of range shifts, should be resolved using the beamlets measured individually for all energies. 
The present approach is also applicable to proton and other ion beams and to many other dosimeter types such as semiconductors, scintillators, radiochemical gels, thermoluminescence and radio-photoluminescence dosimeters, etc., as long as they satisfy the requirements for precision, dynamic range, stability, element size and pitch, phantom perturbation, ease of use, cost, etc.
Such products will be especially useful when the dosimetry involves systematic response variation such as quenching, or when the treatment planning involves model-based dose correction such as RBE.\footnote{Patent pending JP2018-177593, 21 September 2018}

\section{Conclusions}

In this proof-of-concept study, we have developed a method to measure physical and biological doses of carbon-ion RT by longitudinal dosimetry of quenched response without using plan data.
This method only involves a general optimization algorithm, a radiobiology model, and experimental beamlet data, and requires no extra corrections.
The dosimetric precision reflected the intrinsic RCF performance, which may generally satisfy QA requirements.
The range mixing within each beamlet limited the deconvolution and caused minor moderation to the reconstructed dose distributions especially at the ends of SOBP.
Theoretically, this approach is applicable to various dosimeters and to proton and ion beams of any delivery method, regardless of quenching or RBE.

\section*{Disclosure of Conflicts of Interest}

The authors have no conflicts of interest to disclose.

\bibliography{text}

\end{document}